\documentclass[12pt]{article}
\pdfoutput=1
\usepackage{graphicx}
\usepackage{amsmath,latexsym}
\usepackage[small]{subfigure}

\newcommand\pubnumber{UWTHPH 2014-32}
\newcommand\pubdate{\today}

\def\univie{$^*$\,University of Vienna, Faculty of Physics, Boltzmanngasse 5, A-1090 Wien, Austria}
\def\esi{$^\dagger$\,Erwin Schr\"odinger International Institute for Mathematical
Physics, University of Vienna, Boltzmanngasse 9, A-1090 Vienna, Austria}
\def\support{\footnote{Presenter}}

\def\Title#1{\begin{center} {\Large #1 } \end{center}}
\def\Author#1{\begin{center}{ \sc #1} \end{center}}
\def\Address#1{\begin{center}{ \it #1} \end{center}}

\newcommand\pubblock{\rightline{\begin{tabular}{l} \pubnumber\\
         \pubdate  \end{tabular}}}
\newenvironment{Abstract}{\begin{quotation}  }{\end{quotation}}
\newenvironment{Presented}{\begin{quotation} \begin{center} 
             PRESENTED AT\end{center}\bigskip 
      \begin{center}\begin{large}}{\end{large}\end{center} \end{quotation}}
\def\Acknowledgements{\bigskip  \bigskip \begin{center} \begin{large}
             \bf ACKNOWLEDGEMENTS \end{large}\end{center}}

\def\beq{\begin{equation}}
\def\eeq#1{\label{#1}\end{equation}}
\def\eeqn{\end{equation}}

\def\beqa{\begin{eqnarray}}
\def\eeqa#1{\label{#1}\end{eqnarray}}
\def\eeqan{\end{eqnarray}}

\let\bar=\overbar

\def\Dslash{\not{\hbox{\kern-4pt $D$}}}
\def\dslash{\not{\hbox{\kern-2pt $\del$}}}

\def\msb{{\bar{\ssstyle M \kern -1pt S}}}

\begin{document}
\begin{titlepage}
\pubblock

\vfill
\Title{Charm and Bottom Masses from Sum Rules \\with a Convergence Test}
\vfill
\Author{ Bahman Dehnadi\,$^{*,}$\support, Andr\'e H. Hoang\,$^{*,\dagger}$, Vicent Mateu\,$^*$}
\Address{\univie}
\Address{\esi}
\vfill
\begin{Abstract}
In this talk we discuss results of a new extraction of the {$\overline{\rm MS}$} charm quark mass using relativistic 
QCD sum rules at ${\cal O}(\alpha_s^3)$ based on moments of the vector and the pseudoscalar current correlators and using the available 
experimental measurements from $e^+\,e^-$ collisions and lattice results, respectively. The analysis of the perturbative uncertainties is
based on different implementations of the perturbative series and on independent variations of the renormalization scales for the mass and 
the strong coupling following a work we carried out earlier. Accounting for the perturbative series that result from this double scale variation
is crucial since some of the series can exhibit extraordinarily small scale dependence, if the two scales are set equal.
The new aspect of the work reported here adresses the problem that double scale variation might also lead to an overestimate 
of the perturbative uncertainties. We supplement the analysis by a convergence test that allows to quantify the overall convergence of 
QCD perturbation theory for each moment and to discard series that are artificially spoiled by specific choices of the renormalization scales.
We also apply the new method to an extraction of the {$\overline{\rm MS}$} bottom quark mass using experimental
moments that account for a modeling uncertainty associated to the continuum region where no experimental data is available. 
We obtain $\overline m_c(\overline m_c) = \,1.287\,\pm 0.020\,$GeV and $\overline m_b(\overline m_b) = \,4.167\, \pm\, 0.023\,$GeV.
\end{Abstract}
\vfill
\begin{Presented}
International Workshop on the CKM Unitarity Triangle\\
Vienna, Austria,  September 8--12, 2014
\end{Presented}
\vfill
\end{titlepage}
\def\thefootnote{\fnsymbol{footnote}}
\setcounter{footnote}{0}
%
%
{\it Introduction:} Precise and reliable determinations of the charm and bottom quark masses are an important input for a number of theoretical predictions such as Higgs
branching ratios to charm and bottom quarks or for the corresponding Yukawa couplings \cite{Heinemeyer:2013tqa}. They also affect the theoretical predictions of radiative and inclusive B decays,
as well as rare kaon decays. For example, the inclusive semileptonic decay rate of B mesons depends on the fifth power of the bottom quark mass. These weak decays provide crucial methods to determine elements of the CKM matrix 
and which in turn are important for testing the validity of the Standard Model and for indirect searches of new physics. In this context having a reliable estimate of uncertainties of the quark masses is as important as knowing their 
values precisely~\cite{Antonelli:2009ws}.
Due to confinement quark masses are not physical observables. Rather, they are scheme-dependent parameters of the QCD Lagrangian which have to be determined from quantities that strongly depend on them.

One of the most precise tools to determine the charm and bottom quark masses is the QCD sum rule method where weighted averages of the normalized cross section $R_{e^+e^-\to\, q\bar{q}\,+X}$, with $q = c, b$, can be related to moments 
of the quark vector current correlator $\Pi_V$ which can be calculated in perturbation theory using the OPE:
\begin{align}
\label{momentdefvector}
& M_n = \!
\int\!\dfrac{{\rm d}s}{s^{n+1}}R_{e^+e^-\to\, q\bar{q}\,+X}(s)\,,\qquad
R_{e^+e^-\to\, c\bar{c}\,+X}(s) = \dfrac{\sigma_{e^+e^-\to\,
c\bar{c}\,+X}(s)}{\sigma_{e^+e^-\to\,\mu^+\mu^-}(s)}\,, \\
&M_n^{\rm th} =\dfrac{12\pi^2 Q_q^2}{n!}\,\Big(\dfrac{{\rm d}}{{\rm d}q^{2}}\Big)^{n}\Pi_V(q^2)\Big|_{q^2=0}\,,\quad\,
j^{\mu}(x) = \bar{\psi}(x)\gamma^\mu\psi(x)
\,,\nonumber \\
&\big(g_{\mu\nu}\,q^2-q_{\mu}q_{\nu}\big)\Pi_V(q^{2}) 
= -\, i\!\int\!\mathrm{d}x\, e^{iqx}\left\langle \,0\left|T\,
j_{\mu}(x)j_{\nu}(0)\right|0\,\right\rangle \,.\nonumber
\end{align}
Here $Q_q$ is the quark electric charge and $\sqrt{s} = \sqrt{q^2}$ is the $e^+e^-$ center of mass energy. 
Since the integration over the experimental R-ratio stretches from the quark pair threshold up to infinity and since useful experimental measurements only exist for energies up to around 11~GeV, the method relies on using theory input for energies above that scale (which we call ``continuum''). For the charm case the combination of all available measurements is sufficient to render the experimental moments independent of uncertainties one assigns to the theory input for the continuum region~\cite{ourpaper}. For the bottom case the dependence on the continuum theory input is very large, and the dependence of the low-n experimental moments on unavoidable assumptions about the continuum error can be the most important component of the error budget. In fact, the use of the first moment $M_1$ appears to be excluded until experimental data become available for higher energies.

Alternatively one can also consider moments of the pseudoscalar current correlator. Experimental information on the correlator $\Pi_P$ is not available in a form useful for quark mass determinations, but for charm quarks very precise lattice calculations have become available recently~\cite{HPQCD:2008}. For the pseudo-scalar correlator it turns out that the first two Taylor coefficients in the small-$q^2$ expansion need to be renormalized and that the third term (which we will
denote by $M_0^P$) is hardly sensitive to $m_q$. We adopt the definitions
\begin{align}
\label{momentdefpseudo}
&\Pi_P(q^2) 
= i\!\int\!\mathrm{d}x\, e^{iqx}\left\langle \,0\left|T\,
j_P(x)j_P(0)\right|0\,\right\rangle ,\quad j_P(x) = 2\,m_q\,i\,\bar{\psi}(x)\gamma_5\psi(x)\,,\\
&M_n^P =\bigg[\dfrac{12\pi^2}{n!}\,\Big(\dfrac{{\rm d}}{{\rm d}q^{2}}\Big)^{n} P(q^{2})\Big|_{q^2=0}\bigg]^{\frac{1}{2n}}\,, \qquad   
P(q^2) = \dfrac{\Pi_P(q^2) - \Pi_P(0) - \Pi^\prime_P(0)\,q^2 }{q^4}\,,\nonumber
\end{align}
where the explicit mass factor in the definition of the pseudo-scalar current ensures that it is renormalization scheme independent. For the vector as well as the pseudoscalar correlator moments at low values of $n$,
it is mandatory to employ a short-distance mass scheme such as $\overline{\rm MS}$, which renders the quark mass $\overline m_q(\mu_m)$ dependent on its renormalization scale $\mu_m$, similar to the strong coupling $\alpha_s(\mu_\alpha)$ which 
depends on $\mu_\alpha$. 

\begin{figure}[t!]
\includegraphics[width=0.329\textwidth]{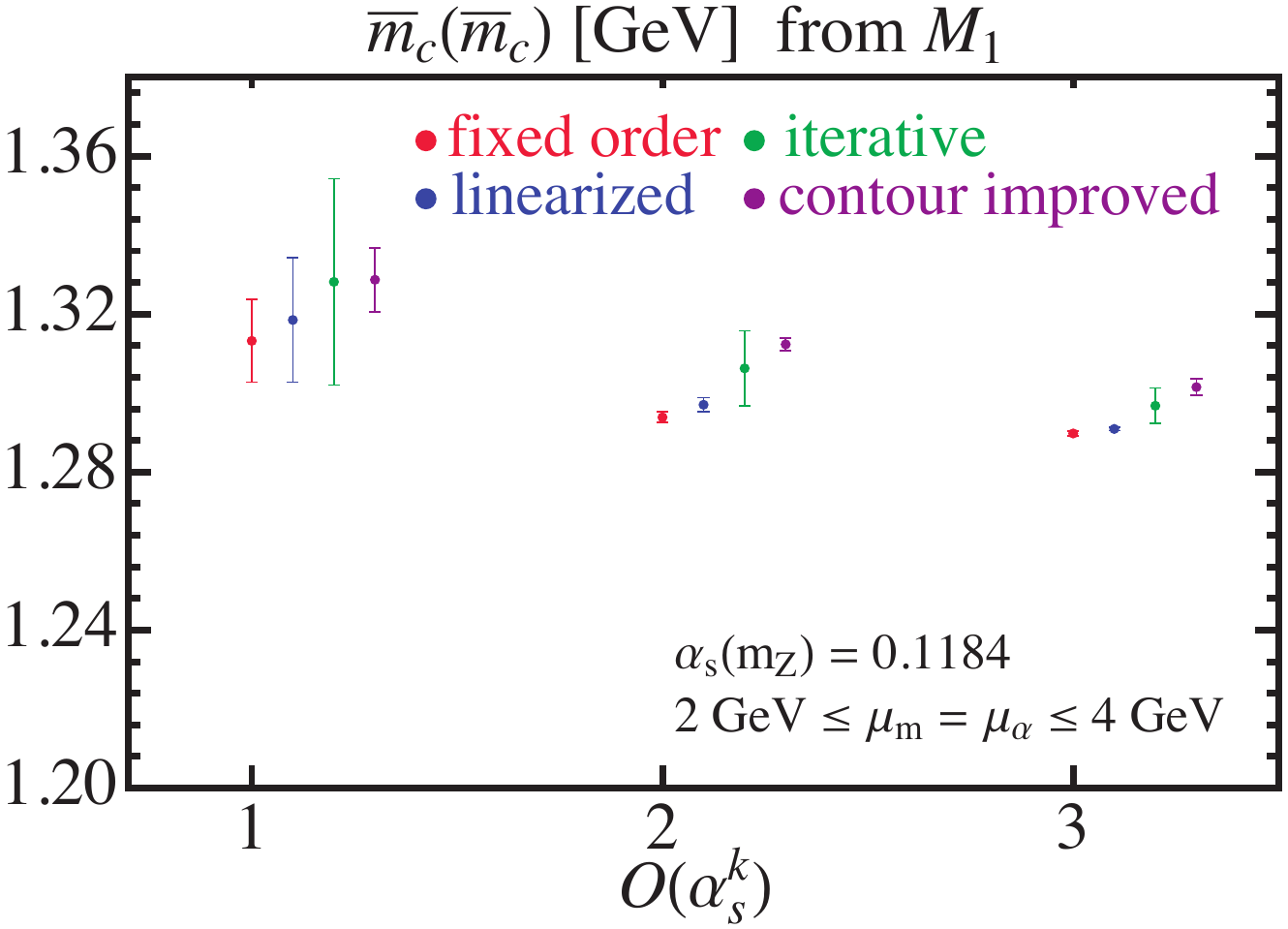}
\includegraphics[width=0.329\textwidth]{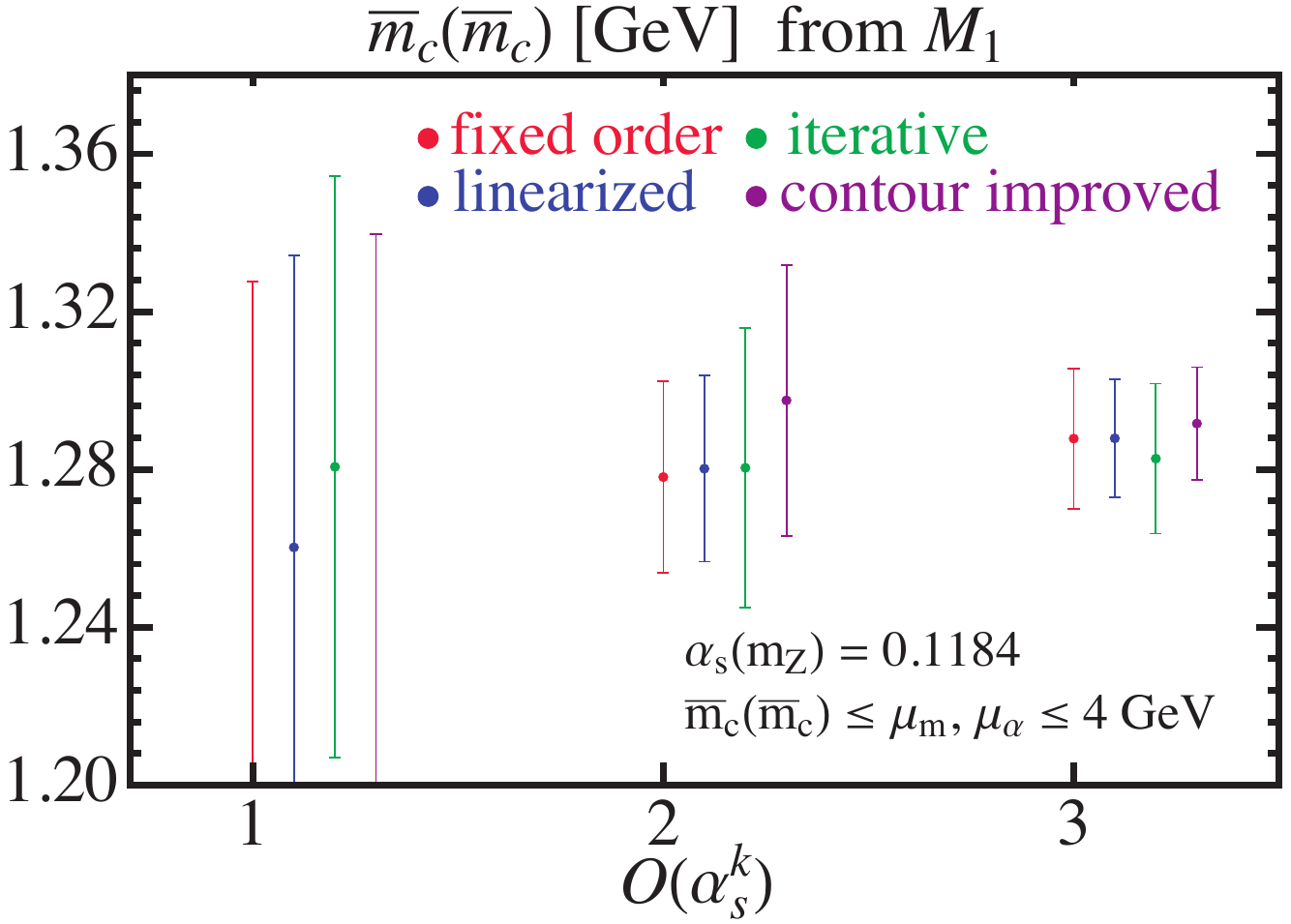}
\includegraphics[width=0.329\textwidth]{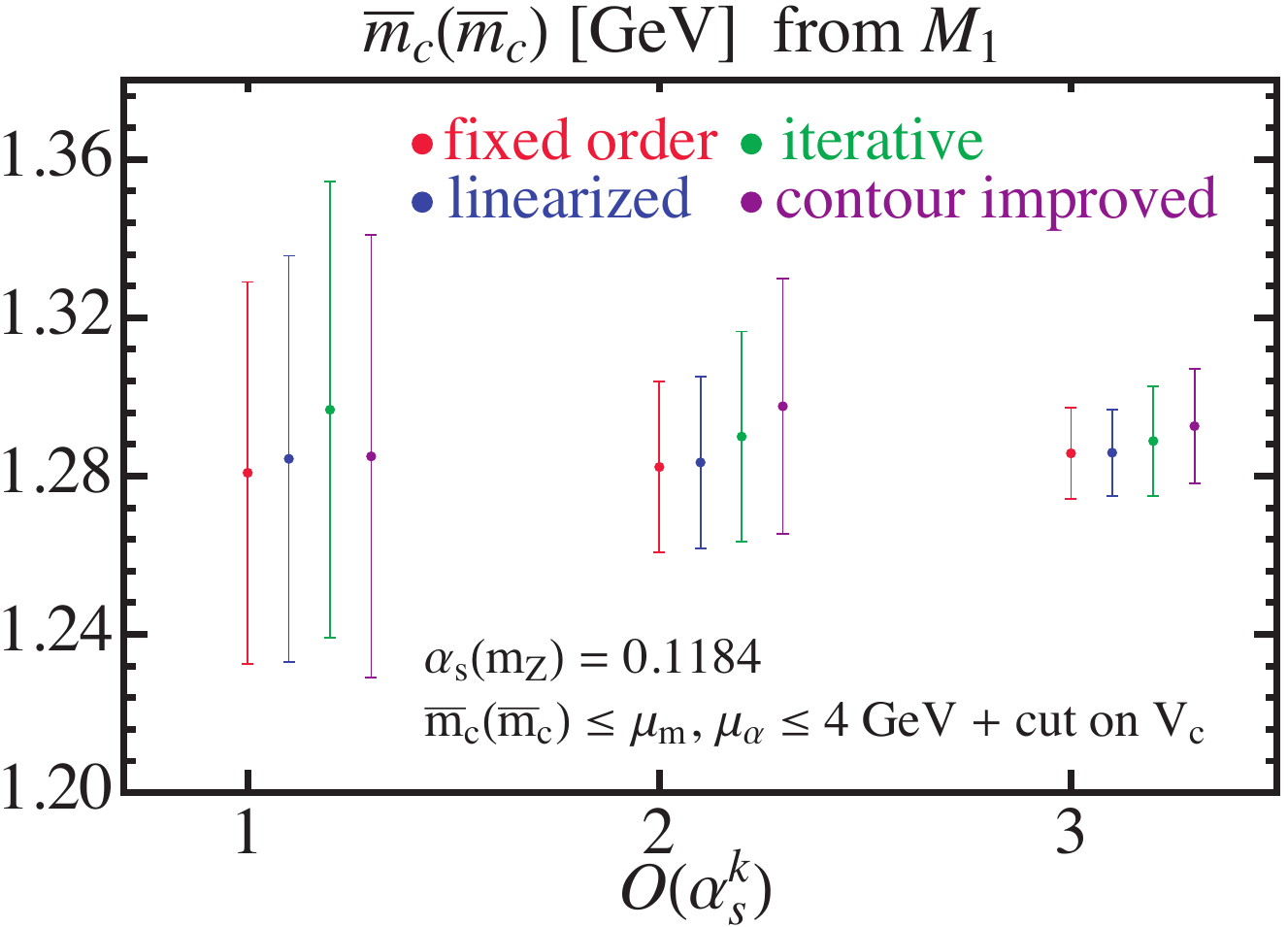}
\vspace*{-0.9cm}
\caption{Charm mass values plus scale variation from the first moment of the vector current at $\mathcal{O}(\alpha_s^{1,2,3})$, for expansions 1-4; (a) correlated scale variation between $2$\,GeV and $4$\,GeV (b) uncorrelated scale variation between and $\overline{m}_c(\overline{m}_c)$ and $4$\,GeV, and (c), as in (b) discarding 3\% of series with highest $V_c$ values.}
\label{fig:order-plots}
\end{figure}

{\it Previous results:} In a number of recent low-$n$ sum rule analyses~\cite{Narison:2012, Bodestein:2011, McNeile:2010, Chetyrkin:2009, HPQCD:2008, Kuhn:2007, Boughezal:2006} based on ${\cal O}(\alpha_s^3)$ perturbation
theory~\cite{Hoang:2008qy, Kiyo:2009, Greyant:2010} the constraint $\mu_m=\mu_\alpha$ was employed. 
In Ref.~\cite{ourpaper} we analyzed the perturbative series for the vector correlator moments $M_{1,2,3,4}$ at ${\cal O}(\alpha_s^3)$ with the aim to reliably assess their uncertainty coming from the truncation of the series and to 
find out whether renormalization scale variation with $\mu_m=\mu_\alpha$ gives a compatible estimate of this uncertainty. We considered four different alternatives, each of which providing a viable analytic expression to carry out the charm mass determination:
\begin{enumerate}
	\item conventional {\bf fixed order} expansion for $M_n$ and solving for $\overline m_q(\mu_m)$ numerically; 
	\item taking the $2n$-th root of $M_n$ and expanding in $\alpha_s(\mu_\alpha)$, which achieves a {\bf linearized} relation of the inverse of the root of the experimental moment to $\overline m_q$;
	\item solving for the mass in a {\bf linearized iterative} way as a function of the experimental moment
	and $\alpha_s(\mu_\alpha)$;
	\item adopting an energy-dependent renormalization scale for $\alpha_s$ of the form
	$\mu_\alpha^2 \to \mu_\alpha^2\,[\,1 - q^2/4\overline{m}_q(\mu_m)^2\,]$ and Taylor-expanding $\Pi_V$ around $q^2=0$. 
\end{enumerate}
The fourth method is related to a {\bf contour improved} dispersive integral for the moments in the complex $q^2$ plane, and the four expansion methods can also be used for the pseudo-scalar correlator. 
The important observations made in Ref.~\cite{ourpaper} were:
\begin{itemize}
	\item Using expansions 1-4 with variations of $\mu_m$ and $\mu_\alpha$ that are correlated (such as $\mu_m=\mu_\alpha$) can lead to charm mass results with very small error estimates that are not compatible with each other. 
	For some expansions it happens that the results from the different orders are incompatible to each other. This indicates that correlated scale variation in general underestimates the perturbative uncertainty.
	\item Uncorrelated, i.e.\ independent variation of $\mu_m$ and $\mu_\alpha$ leads to charm mass results with error estimates that are in general larger, but also fully compatible among the expansions 1-4 and among the different orders.
	\item The size of the perturbative error for the charm mass depends quite strongly on the choice of the lower bound of the range of the scale variation. The choice of the upper bound only has a relatively small impact.
\end{itemize}
Figure~\ref{fig:order-plots}(a) shows the results for $\overline{m}_c(\overline{m}_c)$ for expansions 1-4 and at ${\cal O}(\alpha_s^{1,2,3})$ for $2$~GeV$\le\mu_m=\mu_\alpha\le 4$~GeV. 
For example, in Ref.~\cite{Chetyrkin:2009} and in Ref.~\cite{McNeile:2010} the fixed order and the linearized expansions, respectively, were employed using this type of correlated scale variation, 
leading to a perturbative error estimate at the level of $1$ to $2$~MeV.  
Given these observations we concluded that uncorrelated variations of $\mu_m$ and $\mu_\alpha$ have to be used for a reliable estimate of the perturbative uncertainty. 
We further argued that the proper range of variation should include the charm mass itself, the scale that governs the series, and we adopted the range $\overline{m}_c(\overline{m}_c) \le \mu_m, \mu_\alpha \le 4$~GeV,
which is motivated by the range $2\,m_c\,\pm\, m_c$ around the pair production threshold. The outcome is shown in Fig.~\ref{fig:order-plots}(b) giving a perturbative error estimate of around $20$~MeV. 
The final result quoted in Ref.~\cite{ourpaper} for $\alpha_s(m_Z)=0.1184\pm 0.0021$ was
$\overline m_c(\overline m_c) = 1.282
\, \pm \, (0.006)_{\rm stat} 
\, \pm \, (0.009)_{\rm syst} 
\, \pm \, (0.019)_{\rm pert}
\, \pm \, (0.010)_{\alpha_s} 
\, \pm \, (0.002)_{\langle GG\rangle}\,$GeV based on the expansion 3 (linearized iterative method).

{\it New results with a convergence test:} While certainly being a conservative method, one point of concern in using an uncorrelated scale variation is that it might lead to an overestimate of the perturbative error, e.g.\ due to logarithms $\ln(\mu_m/\mu_\alpha)$ in connection with the low-scale strong coupling $\alpha_s(\mu_\alpha)$ that might artificially spoil the perturbative series. One approach might be to simply reduce the range of scale variation, in particular the lower bound; but this does not resolve the issue since the resulting smaller variation then simply represents a matter of choice. Rather, the issue should be resolved from properties of the perturbative series themselves. In Ref.~\cite{ournewpaper} we address this issue by supplementing the uncorrelated scale variation method with a convergence test constraint, which we explain in the following and which leads to an updated result for the charm mass from the vector current moment analysis. In addition we use the new approach to analyze the charm quark pseudoscalar current moments as well as the vector current moments for bottom quarks. Due to lack of space in these proceedings we only present the main idea and some results and refer to Ref.~\cite{ournewpaper} for details. The convergence test is as follows:
\begin{itemize}
	\item[(a)] For each pair $(\mu_m,\mu_\alpha)$ the convergence parameter $V_c$ is calculated from the 
	charm mass series $\overline{m}_c(\overline{m}_c)=m^{(0)}+\delta m^{(1)}+\delta m^{(2)}+\delta m^{(3)}$ that results from the fits at ${\cal O}(\alpha_s^{0,1,2,3})$: \vspace*{-0.4cm}
	\begin{equation}
	 V_c = \max\!\bigg[\frac{\delta m^{(1)}}{m^{(0)}}\,,\Big(\frac{\delta m^{(2)}}{m^{(0)}}\Big)^{\!\!1/2},\Big(\frac{\delta m^{(3)}}{m^{(0)}}\Big)^{\!\!1/3}\,\bigg].
	\end{equation}
\vspace*{-0.7cm}
	\item[(b)] The resulting distribution of $V_c$ values is a measure for the overall convergence of the perturbative expansion that is employed.
	If the distribution is peaked around the average $\langle V_c\rangle$ it has a well-defined convergence. Hence discarding series with $V_c\gg \langle V_c\rangle$ (which otherwise significantly enlarge the error) is justified.
\end{itemize}
\begin{figure}[t!]
\includegraphics[width=0.32\textwidth]{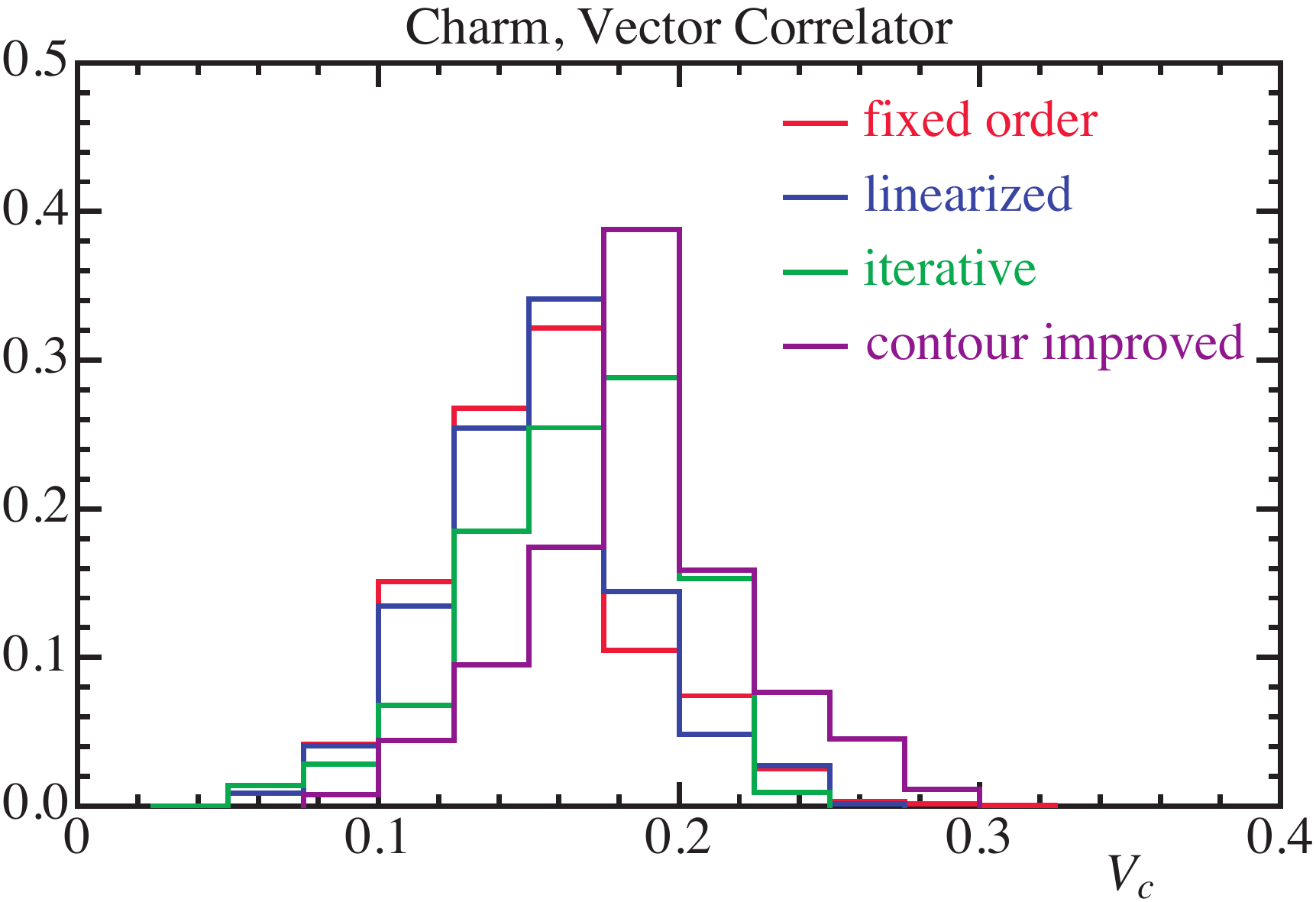}~~
\includegraphics[width=0.32\textwidth]{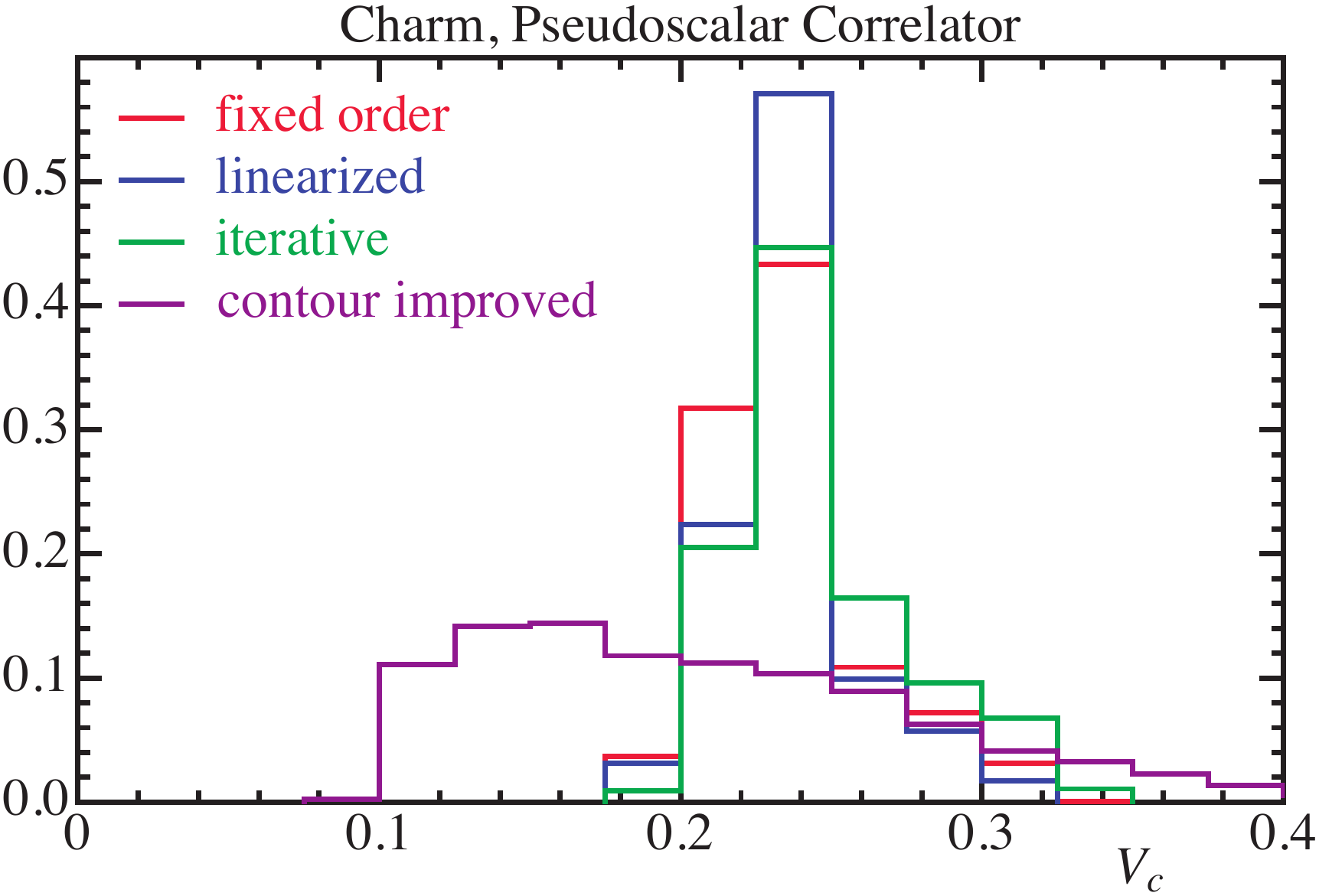}~~
\includegraphics[width=0.31\textwidth]{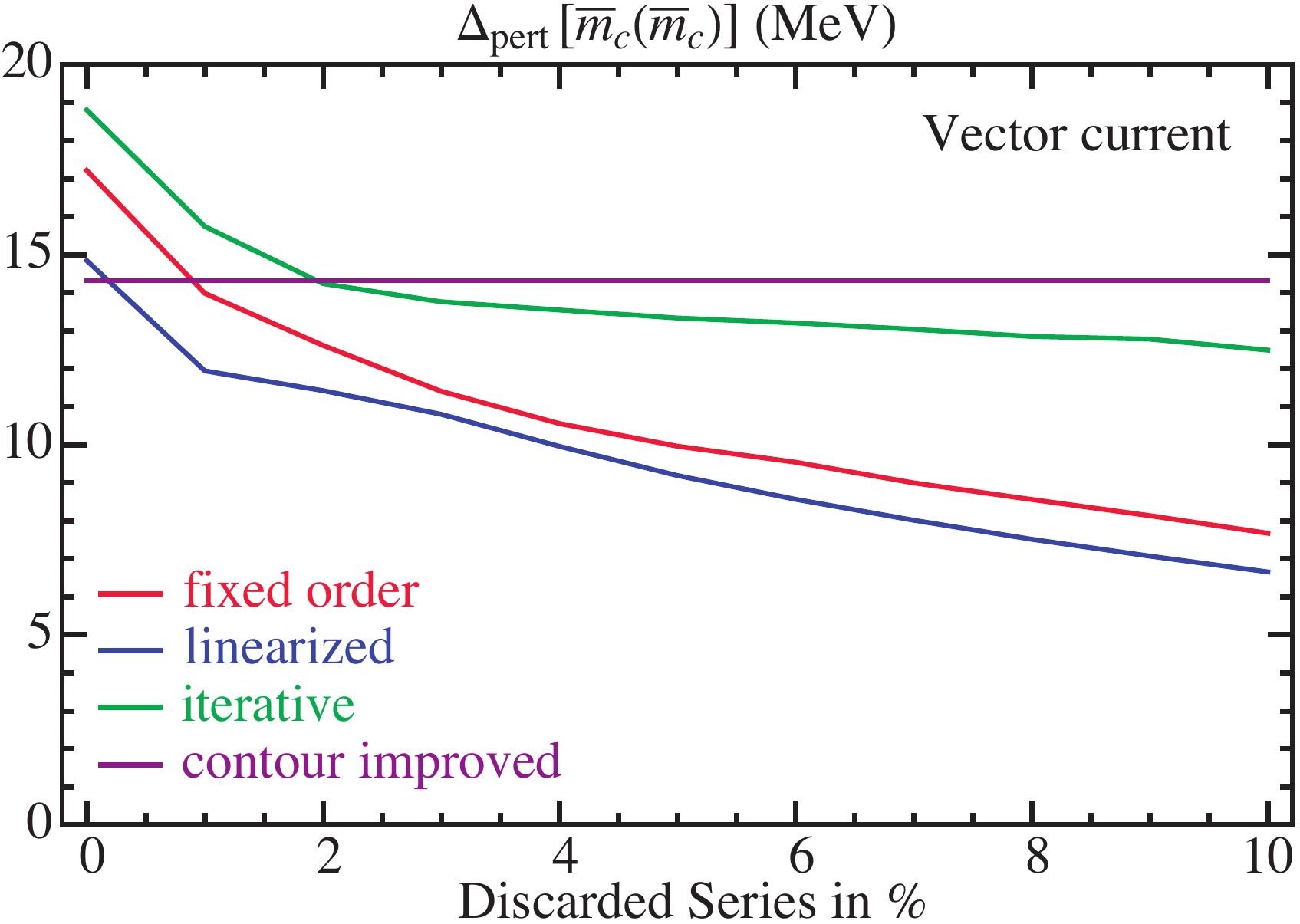}
\vspace*{-0.9cm}
\caption{(a) $V_c$ distribution for $\overline{m}_c(\overline{m}_c)$ from vector $M_1$ for expansions 1-4; (b) same from pseudoscalar $M_1^P$; (c) half of the scale variation on $\overline{m}_c(\overline{m}_c)$ from $M_1$ at ${\mathcal{O}(\alpha_s^3)}$ as a function over the fraction of the discarded series with highest $V_c$ values.}
\label{fig:triming}
\end{figure}
The resulting $V_c$ distributions of expansions (1,\,2,\,3,\,4) for the analysis of the vector moment $M_1$ are shown in Fig.~\ref{fig:triming}a, with $\langle V_c\rangle=(0.15,\,0.15,\,0.17,\,0.19)$, clearly indicating a very good overall convergence. Figure~\ref{fig:triming}c shows the scale variation error (=~half the overall variation) as a function of the fraction of the series (with the largest $V_c$ values) that are being discarded. We see that, indeed, only about 2\% of the series with the highest $V_c$ values are responsible for increasing the scale variation from well below $15$~MeV to up to $20$~MeV. For our final estimate we discard 3\% of the series which leads to an updated result of (using  $M_1^{\rm exp}=0.2121\,\pm\, 0.0020_{\rm stat}\,\pm\, 0.0030_{\rm syst}$~\cite{ourpaper})
\begin{align}\label{eq:vector-result}
\overline m_c(\overline m_c) = & \,1.287
\, \pm \, (0.006)_{\rm stat}
\, \pm \, (0.009)_{\rm syst}
\, \pm \, (0.014)_{\rm pert}\\
&\, \pm \, (0.010)_{\alpha_s}
\, \pm \, (0.002)_{\langle GG\rangle}~{\rm GeV}\nonumber
\end{align}
for the charm mass from the vector correlator analysis using the iterative expansion as the default. Fig.~\ref{fig:order-plots}(c) shows the outcome for the other expansions. We also indicate in Eq.~(\ref{eq:vector-result}) the uncertainty from $\alpha_s(m_Z)\,=\,0.1184\,\pm\,0.0021$ and the gluon condensate. This is the main result of our analysis for the charm mass. We also applied the same approach to extract the charm mass from the pseudo-scalar moment $M_1^P$. The resulting $V_c$ distributions are shown in Fig.~\ref{fig:triming}b and again show a clearly visible peak. However, with $\langle V_c\rangle=(0.24,\,0.24,\,0.25,\,0.21)$, the average $V_c$ values are clearly larger indicating that the pseudoscalar moment has a perturbative convergence that is worse than for the vector moment. This means that the vector correlator method is superior, and that it is expected that the perturbative uncertainty in the charm mass from the pseudo-scalar is larger. We find (using $M_1^{P,{\rm latt}}=0.5127\pm 0.0037$~\cite{HPQCD:2008})
\begin{align}
\overline m_c(\overline m_c) = & \,1.266
\, \pm \, (0.008)_{\rm lat}
\, \pm \, (0.035)_{\rm pert}
\, \pm \, (0.019)_{\alpha_s}
\, \pm \, (0.002)_{\langle GG\rangle}~{\rm GeV}
\end{align}
using again the iterative expansion as the default.

%
%

Using our method we can also determine the  {$\overline{\rm MS}$} bottom quark mass from the vector correlator. For the determination of the experimental moments from the region above $11.2$~GeV we use pQCD (which has essentially negligible errors) supplemented by a modeling uncertainty. Comparing pQCD and rebinned data in the region between $11.06\,$GeV and $11.2\,$GeV we find a $4\%$ discrepancy. Given that the rel.\ discrepancy between experiment and pQCD for $R_b$ at the Z-pole is about $3$ permille~\cite{ALEPH:2010aa}, we adopt a rel.\ modeling error that decreases linearly from $4\% $ at $11.2$~GeV to $3$~permille at $m_Z$, and which is the pQCD error above. This uncertainty makes up for $96.9\%$ of the total error for the first moment $M_1$ (which has an overall $2.45\%$ relative error), and $86.27\%$ of the second moment $M_2$ (which has a overall $1.85\%$ relative error). Note that if we would adopt a constant $4\%$ error for all energies above $11.2$~GeV, this continuum uncertainty would make up for $97.25\%$ of the total error for the first moment $M_1$ (which has an overall $2.59\%$ relative error), and $86.57\%$ of the second moment $M_2$ (which has a overall $1.86\%$ relative error). The difference is small because contributions from higher energies are suppressed. For more details on these considerations we refer to Ref.~\cite{ournewpaper}. In our analysis we use the second moment $M_2$ and employ uncorrelated scale variations in the range $\overline m_b(\overline m_b) \leq \mu_m , \mu_\alpha \leq 15$\,GeV. Interestingly we find that the convergence test for $M_2$ gives $\langle V_c\rangle=(0.13,\,0.11,\,0.12,\,0.15)$ for expansions (1,\,2,\,3,\,4), and we further find that 
discarding series with the highest $V_c$ values only has minor effects on the perturbative error estimate for fractions up to $5\%$. This indicates that the perturbative series for bottom moments are more stable, which is expected from the fact that perturbation theory should work better for the bottom than for the lighter charm. Our final result for the  {$\overline{\rm MS}$} bottom quark mass reads discarding again 3\% of the series with highest $V_c$ values.
\begin{align}
\overline m_b(\overline m_b) = & \,4.167
\, \pm \, (0.004)_{\rm stat}
\, \pm \, (0.018)_{\rm syst}
\, \pm \, (0.010)_{\rm pert}\\
&\, \pm \, (0.007)_{\alpha_s}
\, \pm \, (0.001)_{\langle GG\rangle}~{\rm GeV}\nonumber
\end{align}

\Acknowledgements
We thank the {\it Erwin Schr\"odinger International Institute for Mathematical Physics} (ESI Vienna), where a part of this work has been
accomplished, for partial support. B.D. thanks the FWF Doktoratskollegs ``Particles and Interactions'' (DK W 1252-N27) for partial support.

\end{document}